\documentclass[aps,twocolumn,showpacs,preprintnumbers,amssymb,eqsecnum]{revtex4}
\usepackage{graphicx}
\begin{document}
\title
{Power laws and stretched exponentials 
in a noisy finite-time-singularity model}
\author{Hans C. Fogedby}
\email{fogedby@ifa.au.dk}
\affiliation
{
\footnote{Permanent address}Institute of Physics and Astronomy,
University of Aarhus, DK-8000, Aarhus C, Denmark\\
and\\
NORDITA, Blegdamsvej 17, DK-2100, Copenhagen {\O}, Denmark
}
\author{Vakhtang Poutkaradze}
\email{putkarad@math.unm.edu}
\affiliation
{
Dept of Mathematics and Statistics\\
University of New Mexico\\ 
Albuquerque NM 87131-1141, US
}
\begin{abstract}
We discuss the influence of white noise on a generic
dynamical finite-time-singularity model for a single
degree of freedom. We find that the noise effectively
resolves the finite-time-singularity and replaces
it by a first-passage-time or absorbing state distribution  
with a peak at the singularity and a long time
tail exhibiting power law 
or stretched exponential behavior. The study might
be of relevance in the context of hydrodynamics on
a nanometer scale, in material physics, and in biophysics.
\end{abstract}
\maketitle

\section{\label{secintro}Introduction}
There is a continuing current interest in the influence of noise on
the behavior of nonlinear dynamical systems \cite{Freidlin84,Roy93}.
Here the issues for example associated with the
interference of stochastic noise with the deterministic
chaos of maps \cite{Cvitanovic99} or extended systems, as
for example the noise-driven Kuramoto-Shivashinski
equation \cite{Drotar99}, are of fundamental interest.

In a particular class of systems the nonlinear
character gives rise to finite-time-singularities,
that is solutions which cease to be valid beyond a
particular finite time span. One encounters finite-time-singularities
in stellar structure, turbulent flow, and bacterial growth 
\cite{Kerr99,Brenner97,Brenner98} as well as in 
in Euler flows and in free-surface-flows
\cite{Cohen99,Eggers97,Henderson97}. Finite-time-singularities
are also encountered in modeling in econophysics, geophysics, and
material physics
\cite{Sornette02,Ide01,Sornette01a,Sornette01b,Gluzman01a,Gluzman01b}.

In the context of hydrodynamical flow on a nanoscale \cite{Eggers02},
where microscopic degrees of freedom come into play,
it is a relevant issue how noise influences
the hydrodynamical behavior near a finite time singularity.
Leaving aside the issue of the detailed reduction of the 
hydrodynamical equations
to a nanoscale and the influence of noise on this scale to further
study, we assume in the present context that a single variable or
``reaction coordinate'' effectively captures the interplay between 
the singularity and the noise. 

We thus propose to  
consider a simple generic
model system with  one degree of freedom
governed by 
a nonlinear Langevin equation
driven by Gaussian white noise,
\begin{eqnarray}
\frac{dx}{dt} = - \frac{\lambda}{2|x|^{1+\mu}} + \eta
~,~~~
\langle\eta\eta\rangle(t) = \Delta\delta(t) ~.
\label{lan}
\end{eqnarray}
The model is characterized by the coupling parameter $\lambda$,
determining the amplitude of the singular term, the
index $\mu\geq 0$, characterizing  the nature of the singularity, and the 
noise parameter $\Delta$ determining the strength of the noise correlations.
Specifically, in the case of a thermal environment at temperature
$T$ the noise strength
$\Delta\propto T$.

In the absence of noise this model exhibits a finite-time
singularity at a time $t_0$, where the variable
$x$ vanishes with a power law behavior determined by $\mu$.
When noise is added  the finite-time-singularity event
at $t_0$ becomes a statistical event and is conveniently
characterized by a first-passage-time distribution $W(t)$
\cite{Redner01}.
For zero noise we thus have $W(t)=\delta(t-t_0)$, restating
the presence of the finite-time-singularity. In the presence of noise
$W(t)$ develops a peak about $t=t_0$, vanishes at short times,
and acquires a long time tail.

The model in Eq. (\ref{lan}) has also been studied in the context of
persistence distributions related to the nonequilibrium
critical dynamics of the two-dimensional XY model \cite{Bray00}
and in the context of non-Gaussian Markov processes \cite{Farago00}.
Finally, regularized for small
$x$,  the model enters in connection with an analysis of long-range 
correlated stationary processes
\cite{Lillo02}.  

It follows from our analysis below that for  $\mu=0$,
the logarithmic case,  
the distribution at long times is given by the power
law behavior 
\begin{eqnarray}
W(t)\sim t^{-\alpha}~~,~~~\alpha=\frac{3}{2}+\frac{\lambda}{2\Delta}~.
\label{pow}
\end{eqnarray}
For vanishing nonlinearity, i.e., $\lambda=0$,
the finite-time-singularity is absent and the Langevin equation
(\ref{lan})
describes a simple random walk of the reaction coordinate, yielding the
well-known exponent $\alpha=3/2$ \cite{Redner01,Stratonovich63,Risken89}. 
In the nonlinear case with a finite-time-singularity
the exponent  attains a nonuniversal correction depending on
the ratio of the nonlinear strength to the strength of the noise;
for a thermal environment the correction is proportional to $1/T$.

In the generic case for $\mu>0$ considered here we find that the fall
off is slower and that the correction
to the random walk result is given by a stretched exponential
\begin{eqnarray}
W(t)\sim t^{-3/2}\exp[A(t^{-\mu/(2+\mu)}-1)]~,
\label{stret}
\end{eqnarray}
where $A\rightarrow\lambda/\Delta\mu$ for $\mu\rightarrow 0$;
in the limit $\mu\rightarrow 0$ this expression reduces to 
expression (\ref{pow}). The above results for $\mu=0$ and $\mu>0$
have also been obtained in the context of the critical dynamics
of the XY model \cite{Bray00}.

The paper is organized in the following manner. In Section \ref{secfinite} 
we introduce the 
finite-time-singularity model in the noiseless case and
discuss its properties. In Section \ref{seclangevin} we consider the noisy
case and discuss the ensuing Langevin equation and associated
Fokker Plank equation. We discuss the relationship to an absorbing
state problem and  introduce the
first-passage-time or absorbing state distribution.
In Section \ref{secwkb} we review the  weak noise WKB phase space approach to the
Fokker-Planck equation, apply it to the
finite-time-singularity problem, and discuss
the associated dynamical phase space problem.
In Section \ref{secrandom} we apply the WKB phase space approach and evaluate
the weak noise absorbing state distribution at long times. 
We derive the
random walk result in the linear case for $\lambda=0$, the
power law tail for $\mu=0$, and the stretched exponential behavior
for $\mu>0$. 
In Section \ref{secfokker} we derive an exact solution of the Fokker-Planck
equation in the case $\mu=0$ in terms of a Bessel function
and present an  expression for the 
absorbing state distribution. 
In Section \ref{secsummary} we present a summary and a conclusion.
Details of the phase space method is discussed in Appendix \ref{app1};
aspects of the exact solution in Appendix \ref{app2}.
\section{\label{secfinite}Finite-time-singularity model}
\begin{figure}
\includegraphics[width=\hsize]{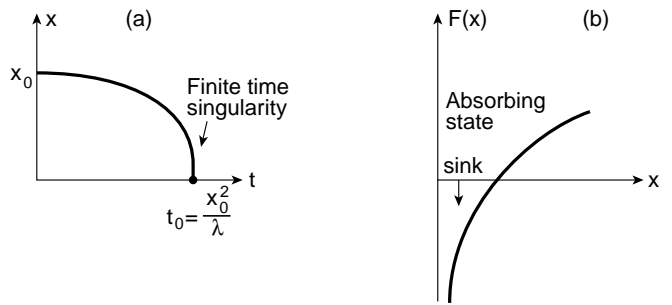}
\caption{
In a) we show the time evolution of the single degree of freedom
$x$. $x$ reaches the absorbing state $x=0$ at a finite time
$t_0$. In b) we depict the free energy $F(x)$ driving the equation.
The absorbing state $x=0$ corresponds to the sink in 
$F(x)$.
}
\label{fig1}
\end{figure}

Let us first consider the noiseless case for $\Delta=0$.
It is instructive to express 
the equation of
motion (\ref{lan}) in the form
\begin{eqnarray}
\frac{dx}{dt} = - \frac{1}{2}\frac{dF}{dx} ~,
\label{eq}
\end{eqnarray}
where the potential or free energy has the form,
\begin{eqnarray}
&&F(x) =  \lambda\log|x|~~~\text{for}~~~\mu=0~,
\label{log}
\\
&&F(x) =  -\frac{\lambda}{\mu}|x|^{-\mu} ~~~\text{for}~~~\mu>0~.
\label{gen}
\end{eqnarray}
The free energy has a logarithmic sink for 
$\mu=0$ and a power law sink for general $\mu>0$.
In both cases $F$
drives  $x$ to the absorbing state $x=0$.
Solving Eq. (\ref{eq}) in the logarithmic case 
we obtain for positive $x$ the solution
\begin{eqnarray}
x=\sqrt\lambda\sqrt{t_0-t}~.
\label{sollog}
\end{eqnarray}
This solution displays a finite-time-singularity at 
$t_0=x_0^2/\lambda$, where $x_0$ is the initial value at time $t=0$,
with  $x$ approaching the absorbing
state with exponent $1/2$. In other words, the attraction to the
sink in the free energy 
occurs in a finite time span; for
times beyond $t_0$ Eq. (\ref{eq}) does not possess a real solution.
This is the way we define a finite-time-singularity in the present
context. 
For general $\mu>0$ we obtain the generalization of the solution
(\ref{sollog}),
\begin{eqnarray}
x=[\lambda(2+\mu)/2]^{1/(2+\mu)}
[t_0-t]^{1/(2+\mu)}~,
\label{solgen}
\end{eqnarray}
which approach the absorbing state
with exponent $1/(2+\mu)$ at time 
$t_0=2x_0^{2+\mu}/\lambda(2+\mu)$.
In Fig.~\ref{fig1} we have for $\mu=0$ depicted the solution $x$ and
the free energy $F(x)$ driving $x$ to zero
\footnote
{
For $\mu<0$ the free
energy vanishes at $x=0$. For $\mu<-2$ $x$ possesses a run-away
solution at a finite time. In the present context we confine 
our discussion to the case $\mu>0$, where the free energy has a sink 
and $x$ approaches an absorbing state.
}
.
\section{\label{seclangevin} Langevin and  Fokker-Planck equations}
In the presence of noise the finite-time-singularity problem
is characterized by the Langevin equation (\ref{lan}).
Here the noise drives the variable $x$ into a fluctuating state
in competition with the free energy which tends to drive $x$ to
the absorbing state $x=0$.  
In the case $\mu\geq 0$ treated here with an absorbing state the
free energy has a sink and there is no stationary distribution;
the probability leaks out at  $x=0$
\footnote
{
In the case $\mu=-2$ the free energy, 
$F=(\lambda/2)x^2$, has the form 
of a harmonic potential and the resulting Langevin equation,
$dx/dt=-(\lambda/2)x+\eta$, governs the stochastic 
dynamics of an overdamped noise-driven harmonic
oscillator. The stationary distribution is given by the Boltzmann
factor $P_0\propto\exp(-F/T)$, where the effective temperature
according to the fluctuation-dissipation theorem is $T=\Delta$. For
$-2<\mu<0$ the same result holds with the free energy given by
Eq. (\ref{gen}).
}
.

From another point of view, introducing the variable $y=x^{2+\mu}$ the Langevin
equation (\ref{lan}) takes the form
\begin{eqnarray}
\frac{1}{2+\mu}\frac{dy}{dt}=-\frac{\lambda}{2}+y^{1/(2+\mu)}\eta~.
\label{abs}
\end{eqnarray}
In the noiseless case the variable $y$  decreases linearly to zero at time 
$t=t_0$.
In the presence of noise 
$y$ fluctuates. We note, however, that the
noise is manifestly quenched at $y=0$, yielding the  
absorbing state.
Absorbing state models of the type in Eq. (\ref{abs})
for extended systems have been studied extensively in the context 
of directed percolation, catalysis, and Reggeon field theory
\cite{Hinrichsen00,Grassberger79,Grassberger82,Marro99}.

In order to analyze the stochastic aspects of finite-time-singularities
in the presence of 
noise we need the time dependent probability distribution
$P(x,t)$
and 
the derived first-passage-time
or absorbing state probability distribution $W(t)$.
The distribution $P(x,t)$ is defined according to 
\cite{vanKampen92,Risken89}
\begin{eqnarray}
P(y,t)=\langle\delta(y-x(t))\rangle~,
\label{dis}
\end{eqnarray}
where $x$ is a stochastic solution of Eq. (\ref{lan}) and 
$\langle\cdots\rangle$ indicates an average over the noise $\eta$
driving $x$. In the absence of noise $P(y,t)=\delta(y-x(t))$, where
$x$ is the deterministic solution of Eq. (\ref{eq}) given by
Eqs. (\ref{sollog}) and (\ref{solgen}) and  depicted in
Fig~\ref{fig1}. At time $t=0$ the variable $x$ evolves 
from the initial condition  $x_0$, implying the boundary condition
\begin{eqnarray}
P(x,0)=\delta(x-x_0)~.
\label{bou}
\end{eqnarray}
At short times  $x$  is close to $x_0$  and the singular term is not yet
operational. In this regime we then obtain ordinary random walk 
with the Gaussian
distribution
\begin{eqnarray}
P(x,t)=(2\pi\Delta t)^{-1/2}
\exp\left[{-\frac{(x-x_0)^2}{2\Delta t}}\right]~,
\label{ran}
\end{eqnarray}
approaching Eq. (\ref{bou}) for $t\rightarrow 0$. At longer times
the barrier $\lambda/2x^{1+\mu}$ comes into play preventing 
$x$ from crossing 
the absorbing state $x=0$. 
This is, however, a random event which can occur at an arbitrary time
instant, i.e., the
finite-time-singularity taking place at $t_0$ in the deterministic
case is effectively resolved in the noisy case. For not too large
noise strength the distribution  is peaked about the noiseless solution
and vanishes for 
$x\rightarrow 0$,  corresponding to the absorbing 
state, implying the boundary condition
\begin{eqnarray}
P(0,t)=0~.
\label{bou2}
\end{eqnarray}
In order to  model a possible experimental situation the first-passage-time
or here absorbing state distribution $W(t)$ is of more direct interest
\cite{vanKampen92,Gardiner97}. First-passage properties
in fact underlie a large class of stochastic processes such as
diffusion limited growth, neuron dynamics, self-organized criticality,
and stochastic resonance \cite{Redner01}.

\begin{figure}
\includegraphics[width=\hsize]{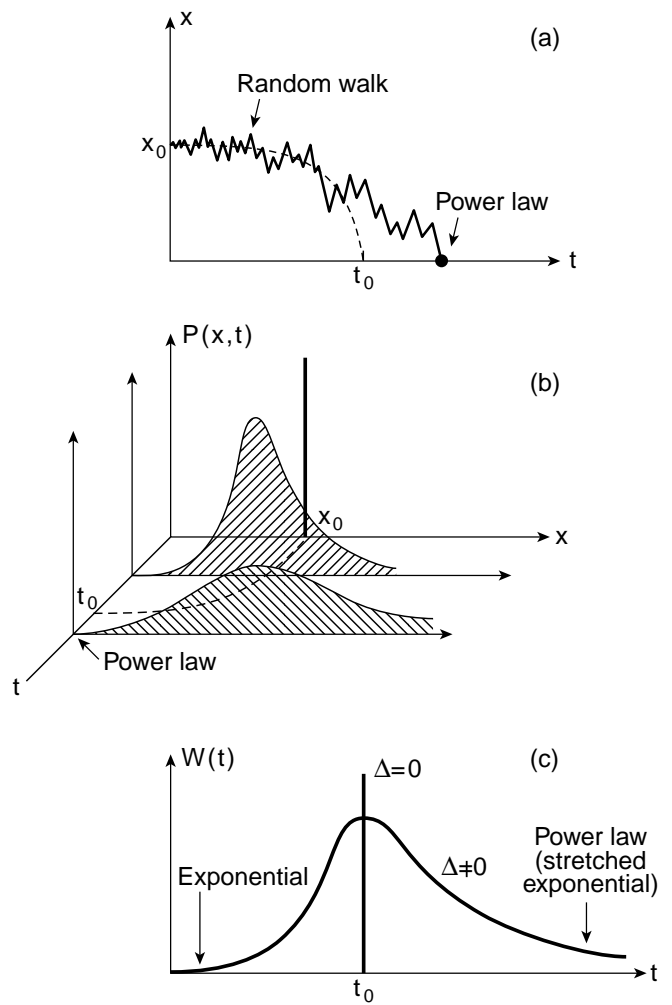}
\caption{
In a) we show a particular  realization of $x$. At early times near
$x_0$ we have random walk behavior. At longer times $x$ is attracted to 
the absorbing state at $x=0$. In b) we depict the distribution $P(x,t)$
in the logarithmic case for $\mu=0$.
For $t=0$ $P(x,0)=\delta(x-x_0)$, i.e., the initial condition. For
larger
$t$ the distribution is broadened about the noiseless trajectory.
$P(x,t)$ exhibits a power law behavior for large $t$ near the absorbing
state. In c)
we show the absorbing state distribution
$W(t)$. For small $t$ the distribution is exponentially small; for large
$t$ it displays a power law behavior for $\mu=0$ and a stretched
exponential behavior in the generic case for $\mu>0$.
}
\label{fig2}
\end{figure}

Since  $P(0,t)=0$ for all $t$  due to the 
absorbing state, the probability that $x$ is not reaching $x=0$
in time $t$ is thus given by $\int_0^\infty P(x,t)dx$, implying  that
the probability
$-dW$ that  $x$ does reach $x=0$ in time $t$ is 
$-dW=-\int_0^\infty dx dt (dP/dt)$. Consequently, the
absorbing state distribution $W(t)$ is determined  by
the expression \cite{Risken89}
\begin{eqnarray}
W(t) =-\int_0^\infty\frac{\partial P(x,t)}{\partial t}dx~.
\label{absd}
\end{eqnarray}
In the absence of noise $P(x,t)=\delta(x-x(t))$ and it follows from
Eq. (\ref{absd}) that $W(t)=\delta(t-t_0)$, 
in accordance with the finite time
singularity at $t=t_0$. For weak noise we anticipate that $W(t)$ will
peak about $t_0$ with vanishing tails for small $t$ and large $t$.
In Fig.~\ref{fig2} we have depicted a particular realization of
$x$ in the noisy case, the distribution $P(x,t)$ in a plot
versus $x$ and $t$,
and the absorbing state distribution $W(t)$.

In the case of Gaussian white noise the distribution $P(x,t)$ satisfies
the Fokker-Planck equation \cite{vanKampen92,Gardiner97}
\begin{eqnarray}
\frac{\partial P}{\partial t} = \frac{1}{2}\frac{\partial}{\partial x}
\left[-\frac{dF}{dx}P+\Delta\frac{\partial P}
{\partial x}
\right]~,
\label{fokker}
\end{eqnarray}
in the present case subject to the boundary conditions 
$P(x,0)=\delta(x-x_0)$ and $P(0,t)=0$.
The absorbing state distribution $W(t)$ then follows from
Eq. (\ref{absd}). 
We note that the Fokker-Planck equation has the form of a conservation
law
$\partial P/\partial t + \partial J/\partial x=0$, defining the 
probability current 
$J=(1/2)(dF/dx)P-(1/2)\Delta\partial P/\partial x$.
Inserting Eq. (\ref{fokker}) in the expression (\ref{absd}) for the 
distribution $W(t)$ and using that $J\rightarrow 0$ for 
$x\rightarrow\infty$ we obtain another expression for $W(t)$:
\begin{eqnarray}
W(t)=\frac{1}{2}\left[\Delta\frac{\partial P}{\partial x}-
P\frac{dF}{dx}\right]_{x=0}~.
\label{absd2}
\end{eqnarray}
The absorbing state distribution is thus equal to the probability
current at the absorbing state.
\section{\label{secwkb}WKB phase space approach}
From a structural point of view the Fokker-Planck equation (\ref{fokker})
has the form of an
imaginary-time Schr\"{o}dinger equation $\Delta\partial P/\partial t=HP$,
driven by the Hamiltonian or Liouvillian $H$. The noise strength
$\Delta$ then plays the role of an effective Planck constant with 
$P$
corresponding to the wavefunction. 
A method  utilizing a non-perturbative WKB phase space approach to a
generic Fokker-Planck equation for extended system was derived in
the context of the Kardar-Parizi-Zhang equation describing
interface growth 
\cite{Fogedby99a,Fogedby99b,Fogedby01c}. In the case of a single
degree of freedom this method amounts to the eikonal approximation
\cite{Risken89,Roy93,Gardiner97}, see also \cite{Graham73,Graham89}. 
For systems with many degrees of 
freedom the method has for example been expounded in \cite{Falkovich96},
based on the functional formulation of the Langevin equation
\cite{Martin73,Janssen76}. In the present formulation 
\cite{Fogedby99a,Fogedby99b,Fogedby01c} the emphasis
is on the canonical phase space analysis
and the use of dynamical system theory \cite{Strogatz94,Ott93}, 
for more details
we refer to Appendix A.  

\begin{figure}
\includegraphics[width=\hsize]{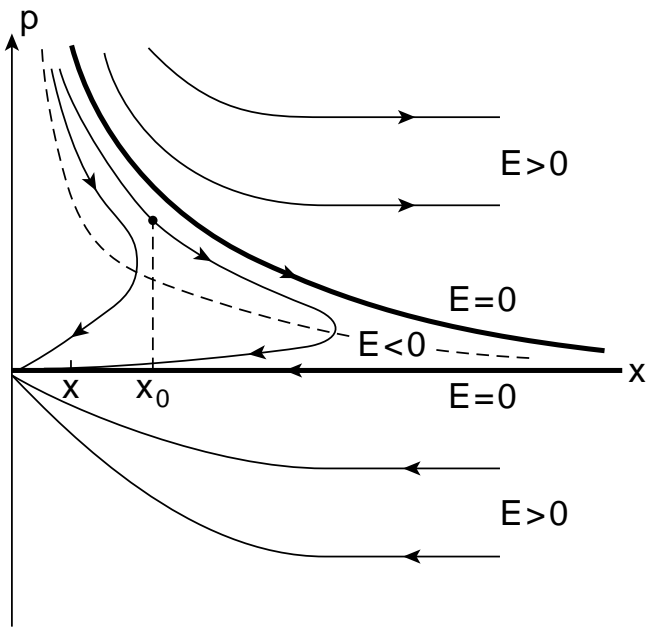}
\caption{
We show the topology of phase space.  The bold lines indicate the zero 
energy submanifolds. We show representative orbits for $E>0$ and $E<0$.
The dashed line indicates a nullcline ($dx/dt=0$) to the 
saddle point $(x,p)=(\infty,0)$.
At long times the orbit from $x_0$ to $x$ migrates to the zero 
energy submanifolds.
}
\label{fig3}
\end{figure}

The weak noise WKB approximation corresponds to the 
ansatz $P\propto\exp[-S/\Delta]$.
The weight function or action $S$ then to
leading asymptotic order in $\Delta$ satisfies a Hamilton-Jacobi
equation $\partial S/\partial t+H=0$ which in turn implies 
a {\em principle of least action}
and Hamiltonian equations of motion
\cite{Landau59b,Goldstein80}.
In the present context the Hamiltonian  has the form
\begin{eqnarray}
H=\frac{1}{2}p^2 - \frac{1}{2}p\frac{dF}{dx}~,
\label{ham}
\end{eqnarray}
yielding the equations of motion
\begin{eqnarray}
&&\frac{dx}{dt}=-\frac{1}{2}\frac{dF}{dx}+p~,
\label{eqx}
\\
&&\frac{dp}{dt}=\frac{1}{2}p\frac{d^2F}{dx^2}~,
\label{eqp}
\end{eqnarray}
replacing the Langevin equation (\ref{lan}). The noise $\eta$
is then represented  by the momentum $p=\partial S/\partial x$ 
conjugate to $x$. The equations (\ref{eqx}) and (\ref{eqp})
determine orbits in a canonical phase space spanned by $x$ and $p$.
Since the system is conserved the orbits lie on the constant
energy manifold(s) given by $E=H$.
The free energy $F$ is given by Eqs. (\ref{log}) and (\ref{gen}) and
the action 
associated with an orbit from $x_0$ to $x$ in time $t$ has the form 
\begin{eqnarray}
S(x_0\rightarrow x,t)=\int_0^t dt\left[p\frac{dx}{dt}-H]\right]~.
\label{action}
\end{eqnarray}
According to the ansatz
the probability distribution is then given by
\begin{eqnarray}
P(x,t)\propto\exp\left[-\frac{S(x_0\rightarrow x,t)}{\Delta}\right]~.
\label{dis2}
\end{eqnarray}
The zero-energy manifold $E=0$ plays an important role 
in determining the long
time distributions. Inserting
$dF/dx=\lambda/x^{1+\mu}$ in Eq. (\ref{ham})
the zero-energy manifold has a submanifold structure given by $p=0$ and
$p=\lambda/x^{1+\mu}$.
According to Eq. (\ref{eqx}) the $p=0$ submanifold corresponds to the
noiseless deterministic motion  given by Eq. (\ref{eq}). In Fig.~\ref{fig3}
we have depicted the canonical phase space spanned by $x$ and $p$.
The heavy lines represent the zero-energy submanifolds $p=0$ and 
$p=\lambda/x^{1+\mu}$. For $E>0$ the energy surfaces are equidistant  for
large $x$ approaching a constant $p$ value; for small $x$ the manifold
$p\sim\lambda/x^{1+\mu}$ for $p>0$ and $p\sim -4Ex^{1+\mu}/\lambda$ for
$p<0$. For $E<0$ the energy surfaces are confined between the 
zero-energy submanifolds; the manifolds approach $(x,p)=(0,0)$ according to
$p\sim 4|E|x^{1+\mu}/\lambda$ and for large $p$ as $p\sim\lambda/x^{1+\mu}$.
For $E \rightarrow -\infty$ the orbits 
approach the positive $p$ half-axis.
The arrows indicate the direction of motion on the manifolds. 
The dashed line indicates a nullcline ($dx/dt=0$) passing through
the hyperbolic fixed point $(x,p)=(\infty,0)$.
In the long time limit the
orbit from $x_0$ to $x$ converges towards the zero-energy
submanifolds.
\section{\label{secrandom}Random walk and long time tails}
The weak noise phase space approach  reviewed above affords a
simple derivation of the asymptotic long time behavior of
the distributions for the finite-time-singularity problem.
In order to derive the transition probability $P(x,t)$
according to Eq. (\ref{dis2}) we simply have to identify
the relevant orbit in phase space from $x_0$ to $x$
which at long times passes close to the zero energy manifolds.
\subsection{\label{subsecrandom}The random walk case}
It is instructive first to consider the case $\lambda=0$. Here the 
finite time singularity at $x=0$ is absent, there is no absorbing
state and the Langevin equation (\ref{lan}) takes
the form $dx/dt=\eta(t)$, describing random walk on the whole axis.
The Hamiltonian is given by $H=(1/2)p^2$ and the equations of 
motion (\ref{eqx}) and (\ref{eqp}) have the form
$dx/dt=p$, $dp/dt=0$ with solutions $p=p_0$ and $x=x_0+p_0t$.
Inserting $dx/dt$ and $H$ in Eq. (\ref{action}) for the action we obtain
$S=(1/2)\int_0^t dt~p^2=(1/2)p_0^2t$ and finally using Eq. (\ref{dis2}) 
the Gaussian distribution (\ref{ran}) for random walk.

\begin{figure}
\includegraphics[width=\hsize]{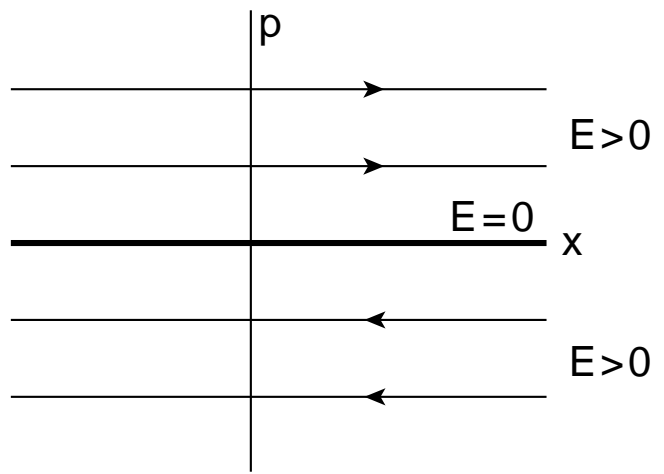}
\caption{
We depict the phase space topology in the random walk case for
$\lambda=0$. The energy manifolds are equidistant corresponding to
constant values of $p$. The bold line for $p=0$ is the 
zero-energy manifold.
}
\label{fig4}
\end{figure}

We note that in order to obtain the correct limit of the
finite-time-singularity problem  we must incorporate the absorbing state
condition at $x=0$. This is achieved by using the method of mirrors
\cite{Risken89} and considering for $P(x,t)$ the linear combination,
\begin{eqnarray}
&&P(x,t)= (2\pi\Delta t)^{-1/2}\times
\nonumber
\\
&&\left(
\exp\left[-\frac{(x-x_0)^2}{2\Delta t}\right] -
\exp\left[-\frac{(x+x_0)^2}{2\Delta t}\right]
\right),
\label{mirror}
\end{eqnarray}
in the half space $x\geq 0$. This distribution is a solution of the
Fokker Planck equation and vanishes for $x=0$. For small $x$
it behaves linearly with $x$,
\begin{eqnarray}
P(x,t)=(2/\pi)^{1/2}
x(\Delta t)^{-3/2}x_0\exp(-x_0^2/2\Delta t)~.
\label{smallx}
\end{eqnarray}
Using Eq. (\ref{absd2}) we readily  obtain the well-known random walk
result
\begin{eqnarray}
W(t)=(2/\pi)^{1/2}
x_0(\Delta t)^{-3/2}\exp(-x_0^2/2\Delta t)~.
\label{absd3}
\end{eqnarray}
For small $t$ the distribution vanishes exponentially. It
displays a  maximum at $t_0=x_0^2/3\Delta$ and falls off algebraically
as $t^{-\alpha}$ for large $t$ with scaling exponent $\alpha=3/2$. 
This behavior is in accordance with the general discussion 
in Section \ref{seclangevin} and is graphically depicted in Fig.~\ref{fig2}.
The phase space topology  in the random walk case is shown in
Fig.~\ref{fig4}.
\subsection{The absorbing state case}
For zero noise and for $\lambda=0$ we have from
Eq. (\ref{lan}) $x=x_0$ at all times, whereas the
solution in the case of a finite-time-singularity
is attracted to the absorbing state at time $t_0$.
In the presence of noise the attraction gives rise
to a change of the form of the absorbing state
distribution $W(t)$ from a delta function peak to a broadened
peak.
For large $\lambda$ the distribution shows
a maximum about $t_0$;  for intermediate values of $\lambda$
the maximum is  between $t_0$ and the random walk value $x_0^2/3\Delta$. 
Because of the attraction to the absorbing state we
also obtain a faster long time fall-off and thus a
positive correction to the random walk exponent $\alpha=3/2$,
depending on the strength $\lambda$.

The plot in Fig.~\ref{fig3} permits a simple qualitative
discussion of the finite-time-singularity phase space
phenomenology. Firstly, a short time orbit from $x_0$ to $x$ 
corresponds to large
value of $|p|$.
In this region and for not too small $x$ the phase space topology
in Fig.~\ref{fig3} is similar to the random walk case depicted
in Fig.~\ref{fig4} and we infer unbiased random walk
behavior, yielding the expression (\ref{mirror}) and (\ref{absd3}).
Secondly, searching for longer time orbits from $x_0$ to  $x$
we must choose smaller $p$ and we move into the region of phase space
for negative energy
where the finite-time-singularity dominates the topology, as 
shown in Fig.~\ref{fig3}. 
In the limit of long times the orbit approaches asymptotically  
the zero-energy submanifolds.
\subsubsection{The logarithmic case  $\mu=0$}
In the logarithmic case  $\mu=0$ the zero energy condition using
Eq. (\ref{ham}) and Eq. (\ref{log}) yields the relationship
$p=\lambda/x$, corresponding
to the hyperbolic manifold; note that the $p=0$ manifold corresponds
to deterministic motion and yields $S=0$. Setting $p=\lambda/x$ and
$H=0$ in the action in 
Eq. (\ref{action}) we then have
$S=\int^t dt~ pdx/dt = \int^x dx~\lambda/x = \lambda\log x$.
Moreover, for $p=\lambda/x$ the equation of motion (\ref{eqx}) reduces
to $dx/dt=\lambda/2x$ with the growing solution
$x^2=\lambda t$.
In the long time limit where the orbit is close to the
zero-energy manifold we thus obtain $S\sim(\lambda/2)\log t$, yielding
according to Eq. (\ref{dis2}) the power law distribution
$P(x,t)\propto t^{-\frac{\lambda}{2\Delta}}$.
Owing to the absorbing state the distribution $P(x,t)$ must
vanish for $x\rightarrow 0$. As discussed in Appendix \ref{app1} this
limit is reproduced by pushing the
WKB approximation in Section \ref{secwkb} to next order in $\Delta$, 
yielding the
correction $-\Delta\log x$ to $S$. Finally, we obtain in the weak noise-
long time limit the distribution
\begin{eqnarray}
P(x,t)\propto xt^{-\frac{\lambda}{2\Delta}}~.
\label{longdis}
\end{eqnarray}
Moreover, applying Eq. (\ref{absd2}) we deduce the weak noise-long time
absorbing state distribution
\begin{eqnarray}
W(t)\propto t^{-\frac{\lambda}{2\Delta}}~,
\label{absdlong}
\end{eqnarray}
with scaling exponent $\alpha=\lambda/2\Delta$
The expressions (\ref{longdis}) and (\ref{absdlong}) 
show that the finite-time-singularity or,
equivalently, absorbing state attracts the random walker 
and increase the fall-off exponent $\alpha$. 
We note that the WKB 
approximation to leading order in $\Delta$ fails to  retrieve the
random walk exponent $3/2$ in the limit $\lambda\rightarrow 0$
and the maximum of $W(t)$ about $t_0$.
\subsubsection{The generic case  $\mu>0$}
In the generic case $\mu>0$ the zero-energy manifold
implies the constraint $p=\lambda/x^{1+\mu}$ and we obtain
similar to the logarithmic case above the action
$S=\int^t dt~pdx/dt=\int^x dx~\lambda/x^{1+\mu}=
-(\lambda/\mu)x^{-\mu} + \text{const.}$
and the equation of motion, $dx/dt=\lambda/2x^{1+\mu}$, 
on the zero-energy manifold with solution
$x^{\mu+2}=(1+\mu/2)\lambda t$. 
In the long time limit we thus find the
action $S=-(\lambda/\mu)((1+\mu/2)\lambda t)^{-\mu/(2+\mu)}
+\lambda/\mu$.
The second order correction to $S$, evaluated in 
Appendix \ref{app1},
is given by $-\Delta(1+\mu)\log x$ and we obtain the 
weak noise-long time distribution
\begin{eqnarray}
&&P(x,t)\propto 
\nonumber
\\
&&x^{1+\mu}
\exp
\left[
\frac{\lambda}{\Delta\mu}
\left[
\left(
\left(
1+\frac{\mu}{2}
\right)
\lambda t
\right)
^{-\frac{\mu}{2+\mu}}
-1
\right]
\right]~,
\label{longdis2}
\end{eqnarray}
and absorbing state distribution
\begin{eqnarray}
W(t)\sim 
\exp
\left[
\frac{\lambda}{\Delta\mu}
\left[
\left(
\left(
1+\frac{\mu}{2}
\right)
\lambda t
\right)
^{-\frac{\mu}{2+\mu}}
-1
\right]
\right]~.
\label{absdlong2}
\end{eqnarray}
The expressions (\ref{longdis2}) and (\ref{absdlong2}) 
show that in the case of a generic
finite-time singularity characterized by the index $\mu$ the power law
behavior of $P(x,t)$ and $W(t)$ is altered to a stretched exponential
behavior depending on $\mu$. In the limit $\mu\rightarrow 0$ we obtain
the power law behavior. We note again that the WKB approximation
is unable to produce the random walk prefactor $t^{-3/2}$ for $\lambda=0$
and the peak of $W(t)$ about $t_0$.
\section{\label{secfokker}Solution of the Fokker-Planck equation}
In this section we return to the Fokker-Planck equation 
(\ref{fokker}) and present
exact expressions for the transition probability $P(x,t)$ and the
absorbing state distribution $W(t)$ in the the logarithmic case
$\mu=0$. We have summarized key points in
the derivation here and defer details to Appendix \ref{app2}.
\subsection{Quantum particle in a repulsive potential}
In the logarithmic case the Fokker-Planck equation assumes the form
\begin{eqnarray}
\frac{\partial P}{\partial t} =
\frac{\Delta}{2}\frac{\partial^2 P}{\partial x^2}
+
\frac{\lambda}{2x}\frac{\partial P}{\partial x}
-
\frac{\lambda}{2x^2}P~.
\label{fokker2}
\end{eqnarray}
Removing the first order term by means of the gauge
transformation $\exp(h)=x^{-\lambda/2\Delta}$
we have
\begin{eqnarray}
-\Delta\frac{\partial}{\partial t}\left[\exp(-h)P\right] =
H\left[\exp(-h)P\right]~,
\label{fokker3}
\end{eqnarray}
where the Hamiltonian $H$ is given by
\begin{eqnarray}
H=-\frac{1}{2}\Delta^2\frac{\partial^2}{\partial x^2}
+
\frac{\lambda^2}{8}\left[1+\frac{2\Delta}{\lambda}\right]
\frac{1}{x^2}~.
\label{ham2}
\end{eqnarray}
This Hamiltonian describes the motion of a unit mass quantum particle 
in one
dimension subject to a centrifugal barrier of strength
$(\lambda^2/8)(1+2\Delta/\lambda)$. For $\lambda=0$ the barrier
is absent and the particle can move over the whole axis; this case
corresponds to ordinary random walk \cite{Risken89}. 
For $\lambda\neq 0$ the particle 
cannot cross the barrier and is confined to either half space; this
corresponds to the case of a finite-time-singularity subject to noise
and an absorbing state at $x=0$.

The Fokker-Planck equation (\ref{fokker3}) has the form of an imaginary
time Schr\"{o}dinger equation with Planck constant $\Delta$ 
for the wavefunction $\exp(-h)P$
and is readily analyzed in terms of Bessel functions \cite{Landau59c}.
Incorporating
the initial condition  $P(x,0)=\delta(x-x_0)$ by defining
$P(x,t)\rightarrow P(x,t)\theta(t)$ we obtain the inhomogeneous differential
equation
\begin{eqnarray}
\frac{\partial P(x,t)}{\partial t} =
\delta(x-x_0)\delta(t)
-\frac{\Delta}{2}e^hHe^{-h}P(x,t)~,
\label{indiff}
\end{eqnarray}
for the determination of $P(x,t)$.
On the positive $x$ and $k$ axis the wavefunctions
$\psi_k(x)=(kx)^{1/2}J_{1/2+\lambda/2\Delta}(kx)$ form, according
to the Fourier-Bessel transform \cite{Mathews73}, 
an orthonormal and complete set satisfying the eigenvalue
equation $H\psi_k(x)=(\Delta^2/2)k^2\psi_k(x)$.
Expanding the right hand side of 
Eq. (\ref{indiff}) on the set $\psi_k$ and using a well-known identity
for Bessel functions \cite{Lebedev72,Gradshteyn65} we obtain for 
the probability distribution  $P(x,t)$
the following closed expression:
\begin{eqnarray}
P(x,t)=\frac
{
x^{\frac{\lambda}{2\Delta}+\frac{1}{2}}
}
{
x_0^{\frac{\lambda}{2\Delta}-\frac{1}{2}}
}
\frac{e^{-\frac{(x^2+x_0^2)}{2\Delta t}}}{\Delta t}
I_{\frac{1}{2}+\frac{\lambda}{2\Delta}}
\left(\frac{xx_0}{\Delta t}\right)~.
\label{exact}
\end{eqnarray}
Here $I_\nu$ is the Bessel function of imaginary argument,
$I_\nu(z)=(-i)^\nu J_\nu(iz)$ \cite{Mathews73}.

By means of Eq. (\ref{absd2}) we moreover deduce the absorbing state
distribution
\begin{eqnarray}
W(t)=\frac{2x_0^{1+\lambda/\Delta}}{\Gamma(1/2+\lambda/2\Delta)}
e^{-x_0^2/2\Delta t}
(2\Delta t)^{-\frac{3}{2}-\frac{\lambda}{2\Delta}}~.
\label{exactabs}
\end{eqnarray}
Similar expressions have also been derived in the
context of the XY model \cite{Bray00}.
\subsection{The distribution $P(x,t)$}
The expression (\ref{exact}) provides the complete solution
of the finite-time-singularity problem for $\mu=0$. The
expression is discussed in more detail in Appendix \ref{app2}.
For $t=0$ we have $P(x,0)=\delta(x-x_0)$ in
accordance with the initial condition (\ref{bou}). For small $t$ we obtain
the random walk result 
$P(x,t)=\exp(-(x-x_0)^2/2\Delta t)/(2\pi\Delta t)^{1/2}$
in accordance with Eq. (\ref{ran}). For $\lambda=0$ we have 
$P(x,t)=(\exp(-(x-x_0)^2/2\Delta t)+x_0\rightarrow -x_0)/
(2\pi\Delta t)^{1/2}$ in agreement with Eq. (\ref{mirror})
for random walk with an absorbing wall at $x=0$.

For long times and $x$ close to the absorbing state $x=0$ we
obtain the asymptotic form
\begin{eqnarray}
P(x,t)\propto
\frac{2xx_0^{1+\frac{\lambda}{\Delta}}e^{-\frac{x^2+x_0^2}{2\Delta t}}}
{\Gamma\left(\frac{3}{2}+\frac{\lambda}{2\Delta}\right)}
(2\Delta t)^{-\frac{3}{2}-\frac{\lambda}{2\Delta}}¨.
\label{longtime}
\end{eqnarray}
For small $x$ the distribution vanishes linearly due to the absorbing
state. For large $t$ the distribution exhibits a power law behavior
with scaling exponent $\alpha=3/2+\lambda/2\Delta$. 
In the weak noise limit the distribution is peaked about the
noiseless solution (\ref{sollog}). For $\lambda=0$ we obtain
the random walk result in Eq. (\ref{smallx}) and we note that the
finite-time-singularity or absorbing state lead to an increase of
the scaling exponent and thus a faster fall-off in time.
In the weak noise limit the scaling exponent approaches $\lambda/2\Delta$
in agreement with the WKB analysis in Section \ref{secwkb}.
In Fig.~\ref{fig2} we have depicted the distribution $P(x,t)$. 
\subsection{The distribution $W(t)$}
The absorbing state distribution $W(t)$ in Eq. (\ref{exactabs})
vanishes exponentially for small $t$. For large $t$ the distribution
shows a power law behavior with exponent $\alpha=3/2+\lambda/2\Delta$. In the
deterministic limit $\Delta=0$ we have $W(t)=\delta(t-t_0)$, where
$t_0=x_0^2/\lambda$. For small $\Delta$ the exponent approaches
$\lambda/2\Delta$ in accordance with Eq. (\ref{absdlong}).
The distribution has a maximum at 
\begin{eqnarray}
t_{\text{max}}=\frac{x_0^2}{3\Delta+\lambda}~.
\end{eqnarray}
For $\Delta=0$ we have $t_{\text{max}}\rightarrow t_0$ and for $\lambda=0$
the random walk result $t_{\text{max}}=x_0/3\Delta$. 
For large coupling strength 
$t_{\text{max}}\rightarrow 0$. Expanding $W(t)$ about $t_{\text{max}}$
we obtain the Gaussian distribution
\begin{eqnarray}
W(t)\propto e^{-(t-t_{\text{max}})^2/\sigma^2}~,
\end{eqnarray}
characterized by the mean square width
\begin{eqnarray}
 \sigma^2=\frac{4\Delta x_0^4}{(3\Delta+\lambda)^3}~.
\end{eqnarray}
Since $W(t)$ falls off as a power of $t$ only a finite number
of moments $\langle t^n\rangle=\int t^n W(t)~dt$ exists.
For  $(2n-1)\Delta<\lambda$ we have
\begin{eqnarray}
\langle t^n\rangle =
\prod_{p=1}^n\left(\frac{x_0^2}{\lambda-(2p-1)\Delta}\right)~.
\label{moments}
\end{eqnarray}
The distribution $W(t)$ is shown in Fig.~\ref{fig2}.
\section{\label{secsummary}Summary and Conclusion}
In this paper we have addressed the problem of the influence
of white Gaussian noise of strength $\Delta$ on a generic
finite-time-singularity of strength $\lambda$, characterized
by the exponent $\mu$. We have for simplicity considered only
a single degree of freedom. We have found that in the case
of a logarithmic sink in the free energy driving the variable,
corresponding to a square root singularity, the first-passage-
time or absorbing state distribution  $W(t)$ displays a peak about the
finite-time-singularity and a long time power law tail 
$\propto t^{-\alpha}$,
characterized by the scaling exponent $\alpha=3/2+\lambda/2\Delta$.
The exponent is nonuniversal and depends on the ratio between
the singularity strength $\lambda$ and the noise strength $\Delta$.
In the case where the noise originates from  a thermal environment
at temperature $T$ we have $\Delta\propto T$ and the scaling
exponent depends on the temperature, $\alpha=3/2+\text{const.}/T$.

In the generic case of a finite-time-singularity characterized
by the exponent $\mu>0$ the weak noise WKB approach shows that
the power law tail for $\mu=0$ is changed to a stretched exponential
with a slower fall-off.

To the extent that the character of a finite-time-singularity
in the vicinity of threshold can be modeled 
with a  single
degree of freedom the present study should hold as regard
the influence of noise on the time distribution. 
We note in particular that in the
case of a thermal environment at temperature $T$ the change
of the scaling exponent becomes large in the limit of
low temperatures as the distribution narrows around
the noiseless threshold time.

The present study also suggests generalizations to the case of 
damping and to the case of several coupled variable subject
to a finite-time-singularity. 

\begin{acknowledgments}
Discussions with D. Diakunov, A. Brandenburg, J. Krug, T. Bohr,
M. H. Jensen, A. Svane and  K. M\o lmer
are gratefully acknowledged. We also thank A. Bray, D. Sornette,
and R. Mantegna 
for drawing our attention to similar results within the
context of the critical dynamics of the XY model,  modeling 
with finite-time-singularities in econophysics and related fields, and
long-range correlated stationary Markov processes, respectively.
\end{acknowledgments}
\appendix
\section{\label{app1}The phase space method}
The weak noise WKB approximation applied to the Fokker-Planck equation
is well-documented \cite{Risken89,Freidlin84,Gardiner97,Graham89}.
Here we review this method with emphasis on a canonical phase space
approach which we have found useful in discussing the pattern formation
and scaling in the noisy Burgers equation 
\cite{Fogedby99a,Fogedby01a,Fogedby01b,Fogedby01c,Fogedby01d}.
We also note that the approach follows from a saddle point approximation
to the functional Martin-Siggia-Rose approach to nonlinear
Langevin equations 
\cite{Martin73,Bausch76,Janssen76,deDominicis78}.
\subsection{To leading order $\Delta$}
Taking as our starting point a generic Langevin equation for one
degree of freedom $x$ driven
by Gaussian white noise,
\begin{eqnarray}
\frac{dx}{dt}=-\frac{1}{2}G(x)+\eta(t)~, ~~\langle\eta\eta\rangle(t)=\delta(t)~,
\end{eqnarray}
the associated Fokker-Planck equation for the distribution $P(x,t)$
takes the form
\begin{eqnarray}
\frac{\partial P}{\partial t}=\frac{1}{2}\frac{\partial}{\partial x}
\left[GP+\Delta\frac{\partial P}{\partial x}\right]~.
\label{fp}
\end{eqnarray}
Applying like in quantum mechanics \cite{Landau59c} the WKB approximation
\begin{eqnarray}
P(x,t)\propto\exp\left[-\frac{S(x,t)}{\Delta}\right] ~,
\label{wkb}
\end{eqnarray}
and expanding the action $S$ in powers of the noise strength
$\Delta$, $S=S_0+\Delta S_1$, $S_0$ satisfies the Hamilton-Jacobi equation
\cite{Landau59b}
\begin{eqnarray}
\frac{\partial S_0}{\partial t}+H(p,x)=0~,~~~~p=\frac{\partial S_0}{\partial x}~,
\label{hj}
\end{eqnarray}
implying a {\em principle of least action} \cite{Landau59b}. 
The action $S_0$ thus has the form
\begin{eqnarray}
S_0(x,t)=\int dt~\left(p\frac{dx}{dt} - H\right)~,
\label{act}
\end{eqnarray}
with Hamiltonian given by
\begin{eqnarray}
H=\frac{1}{2}p(p-G)~,
\label{ham3}
\end{eqnarray}
implying the Hamilton equations of motion
\begin{eqnarray}
&&\frac{dx}{dt}=-\frac{1}{2}G+p~,
\label{hameq1}
\\
&&\frac{dp}{dt}=\frac{1}{2}p\frac{dG}{dx}~.
\label{hameq2}
\end{eqnarray}
The deterministic coupled equations (\ref{hameq1}) and (\ref{hameq2}) replace
the stochastic Langevin equation (\ref{lan}) for weak noise.
The noise $\eta$ is replaced by the canonically conjugate momentum
$p$, which by means of  Eqs. (\ref{wkb}) and (\ref{hj}) is given in terms of
the distribution $P$
\begin{eqnarray}
p=-\Delta\partial\log P/\partial x~.
\label{pint}
\end{eqnarray}

The equations of motion define orbits in a $(x,p)$ phase space
lying on the constant energy surfaces $E=H$ and the general discussion
of the original stochastic problem is replaced by an analysis
of the phase space topology. The prescription
for deriving the distribution to leading order in $\Delta$ thus
amounts to i) solve the equations of motion (\ref{hameq1}) and 
(\ref{hameq2})
for an orbit from an initial value $x_0$ to a final value $x$ reached
in the time span $t$, $p$ being a slaved variable, ii) evaluate the
action $S$ associated with an orbit according to Eq. (\ref{act}), and,
finally, iii) derive  the transition probability from $x_0$ to $x$
in time $t$ using the ansatz (\ref{wkb}). The  zero-energy
manifold here play an important role in determining the
long time distributions. In the limit $t\rightarrow\infty$
a given finite time orbit from $x_0$ to $x$ thus converges
to the zero-energy manifold.

In the case of an overdamped harmonic oscillator described by the
Langevin equation $dx/dt=-\omega x +\eta$ the phase space analysis
was carried out in \cite{Fogedby99a}. In the case of random walk
given by $dx/dt=\eta$ the analysis is performed in 
Section \ref{subsecrandom}. 
\subsubsection{Finite-time-singularity case}
In the generic finite-time-singularity case, $G=\lambda/x^{1+\mu}$,
and we have the Hamiltonian $H=(1/2)p(p-\lambda/x^{1+\mu})$, yielding
the equations of motion $dx/dt=-\lambda/2x^{1+\mu}+p$ and
$dp/dt=-\lambda(1+\mu)p/2x^{2+\mu}$. 
These equations are, however, 
not particularly tractable and we therefore only consider 
the logarithmic case for $\mu=0$. Here $G=\lambda/x$
and we obtain the Hamiltonian
\begin{eqnarray}
H=\frac{1}{2}p\left(p-\frac{\lambda}{x}\right)~,
\end{eqnarray}
and the equations of motion
\begin{eqnarray}
&&\frac{dx}{dt}= -\frac{\lambda}{2x}+p~,
\label{eom1}
\\
&&\frac{dp}{dt}= -\frac{\lambda}{2}\frac{p}{x^2}~.
\label{eom2}
\end{eqnarray}

These equations have a hyperbolic fixed point at
$(x,p)=(\infty,0)$. The nullcline $dx/dt=0$ to the saddle
point is given by $p=\lambda/2x$, indicated in Fig.~\ref{fig3}
by the dashed line. The zero-energy manifolds are given by
$p=0$ and $p=\lambda/x$. The conserved energy $H=E$ provides
the first constant of integration. Solving for $x$ we
have $x=\lambda/(p^2-2E)$. For $E>0$ $p\rightarrow\pm(2E)^{1/2}$
for $x\rightarrow\infty$ as indicated in  Fig.~\ref{fig3}, yielding
in that limit the random walk phase space topology depicted in 
Fig.~\ref{fig4}. For $E<0$ $x=\lambda/(p^2+2|E|)$ exhibiting  a
maximum at the nullcline in a plot of $x$ versus $p$ as shown
for two representative orbits in Fig.~\ref{fig3}. 
Using energy conservation to solve the equation of motion for $p$
and subsequently for $x$ we obtain the solutions
$x^2=(t+t_1)(\lambda+2E(t+t_1))$ and $p^2=2E+\lambda/(t+t_1)$,
where $t_1$ is the second constant of integration.
A specific orbit  from  $x_0$ to $x_1$ in time $t$ thus determines
the constants $E$ and $t_1$; the momentum $p$ becomes
a slaved variable, and the action evaluated along the orbit
yields the distribution.

Considering as final value the absorbing state $x_1=0$, 
the long time orbits lie
in the negative energy region and eliminating $E$ we obtain the solution,
$0<t'<t$
\begin{eqnarray}
x^2=\left(x_0^2\left(1-t'/t\right)+\lambda t'\right)
\left(1-t'/t\right)~.
\end{eqnarray}
The energy is given by $|E|=(\lambda t-x_0^2)/t^2$ and we note that
the energy approaches zero in the long time limit, i.e., the orbit
from $x_0$ to $x=0$ migrates to the zero-energy manifold.
Finally, 
for the action associated with the orbit we obtain 
\begin{eqnarray}
S_0=\frac{1}{2}\left[\lambda\log\frac{\lambda t}{x_0^2}-1\right]~,
\end{eqnarray}
yielding the long time distribution
\begin{eqnarray}
P(x_0\rightarrow 0,t)\propto t^{-\frac{\lambda}{2\Delta}}~,
\end{eqnarray}
in accordance with the expression (\ref{longdis}).

Alternatively, eliminating the momentum $p$ the equations
of motion (\ref{eom1}) and (\ref{eom2}) reduce to a second
order equation for $x$, $d^2x/dt^2=-dV/dx$,
describing the motion of a particle of unit mass in the
attractive potential $V(x)=-(1/8)\lambda^2/x^2$. It then follows
by simple quadrature that all direct orbits to the absorbing state $x=0$ take
a finite time, whereas  the traversal time of negative-energy 
orbits with a turning point diverges in the limit $|E|\rightarrow 0$;
this is in accordance with the phase space behavior shown in 
Fig~\ref{fig3}.

\subsection{Next leading order in $\Delta$}
The next leading order in $\Delta$ is obtained from $S_1$
which by insertion satisfies the equation of motion
\begin{eqnarray}
-\frac{\partial S_1}{\partial t} = 
\left(\frac{\partial S_0}{\partial x}-\frac{G}{2}\right)
\frac{\partial S_1}{\partial x} + 
\frac{1}{2}\frac{d}{dx}
\left(G-\frac{\partial S_0}{\partial x}\right),
\label{next}
\end{eqnarray}
where $\partial S_0/\partial x$ is obtained from the
first order solution.
%
\subsubsection{Random walk case}
From the random walk case discussed in Section \ref{subsecrandom}
we have $S_0=(x-x_0)^2/2t$.
Consequently, Eq. (\ref{next}) takes the form
\begin{eqnarray}
-\frac{\partial S_1}{\partial t} = \left(\frac{x-x_0}{t}\right)
\frac{\partial S_1}{\partial x} -\frac{1}{2}t~,
\label{rwnext}
\end{eqnarray}
with a particular time-dependent solution

\begin{eqnarray}
S_1 = \frac{1}{2}\log|t|~,
\label{rwsol}
\end{eqnarray}
yielding  $S = (x-x_0)^2/t + (\Delta/2)\log|t|$
and the Gaussian distribution
\begin{eqnarray}
P(x,t)\propto |t|^{-1/2}\exp\left[-\frac{(x-x_0)^2}{\Delta t}\right]~.
\label{rwdis}
\end{eqnarray}
As in the quantum case \cite{Landau59c} the next leading correction 
yields the normalization factor $|t|^{-1/2}$.
\subsubsection{Finite-time-singularity case}
In the finite-time-singularity case
$G=\lambda/x^{1+\mu}$ and from above 
$\partial S_0/\partial x=0$. We then obtain inserting in
Eq. (\ref{next})
\begin{eqnarray}
-\frac{\partial S_1}{\partial t}=
-\frac{\lambda}{2x^{1+\mu}}\frac{\partial S_1}{\partial x}
-\frac{1+\mu}{2}\frac{\lambda }{x^{2+\mu}}~,
\end{eqnarray}
with a particular space-dependent solution
\begin{eqnarray}
S_1 = -(1+\mu)\log x ~,
\end{eqnarray}
giving rise to the factor $x^{1+\mu}$ in Eq. (\ref{longdis2}).

\section{\label{app2}The Fokker Planck equation}
Here we discuss the Fokker-Planck equation (\ref{fokker2}) in the
logarithmic case $\mu=0$ in more detail. Applying the gauge
transformation, $\exp(h)=x^{-\lambda/2\Delta}$, and
incorporating the boundary condition (\ref{bou}), $P(x,0)=\delta(x-x_0)$,
we obtain the inhomogeneous differential equation (\ref{indiff}),
with Hamiltonian $H$ given by Eq. (\ref{ham2}) corresponding to the motion
of a  quantum particle subject to a centrifugal barrier
$\propto 1/x^2$.
\subsection{Exact solution}
The right hand side of Eq. (\ref{fokker2}) has the same form as
the standard 
Bessel equation \cite{Lebedev72,Gradshteyn65}. Noting also
the analogy to the quantum case of particle motion in spherical coordinates
\cite{Landau59c} it follows that 
\begin{eqnarray}
\psi_k(x) = (kx)^{1/2}Z_\nu(kx)~,
\label{state}
\end{eqnarray}
where $Z_\nu(kx)$ is a solution of the Bessel equation,
satisfies the eigenvalue equation
\begin{eqnarray}
H\psi_k(x) = k^2\psi_k(x)~, 
\end{eqnarray}
for $\nu = \pm(1/2+\lambda/2\Delta)$. The Bessel function of the
first kind $J_\nu(kx)$ satisfies the absorbing state  boundary condition 
$J_\nu\rightarrow 0$ for $x\rightarrow 0$ and the completeness
and orthogonality of $\psi_k(x)$ follow from the Fourier-Bessel
integral representation \cite{Lebedev72}
\begin{eqnarray}
f(r)=\int_0^\infty dk\int_0^\infty dr'k
J_\nu(kr)J_\nu(kr')f(r')~,
\end{eqnarray}
valid for $\nu>1/2$. We proceed by Fourier transforming (\ref{indiff}), 
\begin{eqnarray}
P(x,t) = \int \frac{d\omega}{2\pi}e^{-i\omega t}p_\omega(x)~,
\end{eqnarray}
and subsequently expanding $p_\omega(x)x^{\lambda/2\Delta}$
and $\delta(x-x_0)$ on the eigenfunctions $\psi_k(x)$,
\begin{eqnarray}
&&p_\omega(x)x^{\lambda/2\Delta}=\int_0^\infty dk\psi_k(x)p_{\omega k}~,
\\
&&\delta(x-x_0)=\int_0^\infty dk\psi_k(x)\psi_k(x_0)~,
\end{eqnarray}
yielding the expansion coefficients
\begin{eqnarray}
p_{\omega k}=
\frac{x_0^{\lambda/2\Delta}\psi_k(x_0)}
{-i\omega + \Delta k^2/2}~.
\end{eqnarray}
Finally, integrating over $\omega$ in the lower half plane
and picking up contributions from the branch cut
we obtain
\begin{eqnarray}
&&P(x,t)=
\nonumber
\\
&&\int_0^\infty dk~ke^{-\Delta k^2t/2}
(xx_0)^{1/2}(x/x_0)^{\lambda/2\Delta}
\times
\nonumber
\\
&&J_{1/2+\lambda/2\Delta}(kx)
J_{1/2+\lambda/2\Delta}(kx_0)~.
\end{eqnarray}
This integral can be reduced further using the identity
\cite{Gradshteyn65}
\begin{eqnarray}
&&\int_0^\infty dx~x e^{-\rho^2x^2}
J_\nu(\alpha x)J_\nu(\beta x)=
\nonumber
\\
&&\frac{1}{2\rho^2}e^{(\alpha^2+\beta^2)/4\rho^2}I_\nu(\alpha\beta/2\rho^2)~, 
\end{eqnarray}
yielding  Eq. (\ref{exact}) in Section \ref{secfokker}.
\subsection{Random walk, short time, and long time limits}
In the random walk case for $\lambda=0$ we obtain using 
$I_{1/2}=2(1/2\pi x)^{1/2}\sinh x$, the expression (\ref{mirror}).
In the short time limit $t\ll xx_0/\Delta$, using 
$I_\nu(x)\propto(1/2\pi x)^{1/2}\exp(x)$ for $x\rightarrow\infty$ 
\cite{Lebedev72} we
obtain  Eq. (\ref{ran}) and for $t=0$ the boundary condition
(\ref{bou}).

In the long time limit $t\gg xx_0/\Delta$ using
$I_\nu(x)\propto(x/2)^\nu/\Gamma(\nu+1)$ for $x\rightarrow 0$
\cite{Lebedev72} we obtain Eq. (\ref{longtime})
Using Eq. (\ref{absd2}) and $\Gamma(z+1)=z\Gamma(z)$ we finally obtain the
absorbing state  distribution (\ref{exactabs}).

The moments of $W(t)$ are easily worked out. Using
$\Gamma(z)=\int_0^\infty t^{z-1} e^{-t}~dt$ \cite{Lebedev72,Gradshteyn65}
we have
\begin{eqnarray}
\langle t^n\rangle = \int t^n W(t)~dt = \left(\frac{x_0^2}{2\Delta}\right)^n
\frac{\Gamma(\frac{1}{2}+\frac{\lambda}{2\Delta} -n)}
{\Gamma(\frac{1}{2}+\frac{\lambda}{2\Delta})}~,
\end{eqnarray}
or further reduced for $(2n-1)\Delta<\lambda$ the expression
(\ref{moments}).
\subsection{Weak noise limit}
In the limit $\Delta\rightarrow 0$ the distribution $P(x,t)$ is centered
about the noiseless solution (\ref{sollog}).
In terms of the
exact solution (\ref{exact}) this is a singular limit since both
order and argument in $I_\nu(x)$ diverge. Using the spectral
representation \cite{Lebedev72,Gradshteyn65}
\begin{eqnarray}
&&I_{\nu}(z)=
\frac{(z/2)^\nu}{\Gamma(\nu+\frac{1}{2})\Gamma(\frac{1}{2})}
\int_0^\pi\cosh(x\cos\theta)\sin^{2\nu}\theta~d\theta~,
\nonumber
\\
&&
\end{eqnarray}
introducing the variable $u$ according to
\begin{eqnarray}
\sinh u = \frac{\lambda t}{2 x x_0}~,
\end{eqnarray}
defining
\begin{eqnarray}
f_\pm(\theta)=\log\sin\theta+\frac{1}{2}
\left(1\pm\frac{\cos\theta}{\sinh u}\right)~,
\end{eqnarray}
and using 
\begin{eqnarray}
\Gamma\left(1+\frac{\lambda}{2\Delta}\right)
\Gamma\left(\frac{1}{2}\right)
\approx\pi\sqrt 2 e^{-\frac{\lambda}{2\Delta}}
\left(\frac{\lambda}{2\Delta}\right)^{\frac{1}{2}+\frac{\lambda}{2\Delta}}~,
\end{eqnarray}
for small $\Delta$ we obtain by insertion in Eq. (\ref{exact})
\begin{eqnarray}
P(x,t)\approx&&\frac{1}{4\pi\sqrt 2}\frac{1}{x_0}\frac{\lambda}{\Delta}
\left(\frac{x_0^2}{\lambda t}\right)^{\frac{1}{2}+\frac{\lambda}{2\Delta}}
e^{-(x^2+x_0^2)/2\Delta t}\times
\nonumber
\\
&&\int_0^\pi d\theta\frac{\sin\theta}{\sinh u}
\left[
e^{\frac{\lambda}{\Delta}f_+(\theta)}+
e^{\frac{\lambda}{\Delta}f_-(\theta)}
\right]~.
\label{asymp}
\end{eqnarray}
The expression (\ref{asymp}) for $P(x,t)$ is directly amenable
to an asymptotic analysis for $\Delta\rightarrow 0$ by means of
Laplace's method \cite{Carrier66}.
For small $\Delta$ the main contributions to the integral originate
from the maxima of $f_+(\theta)$ and $f_-(\theta)$. The two maxima
in the interval $0<\theta<\pi$ are given by
$\cos\theta\pm=\pm\exp(-u)$, yielding 
$f_\pm''(\theta_\pm)=-\coth u$ and 
$f_\pm(\theta_\pm)=(1/2)(\log(1-e^{-2u})+\coth u)$. Performing the
Gaussian integrals about the maxima we thus obtain the asymptotic
result valid for small $\Delta$ and fixed $u$, i.e., fixed $x/t$.
\begin{eqnarray}
&&P(x,t)\approx
\left(
\frac{\lambda}{2\pi\Delta}
\right)^{\frac{1}{2}}
\frac{1}{x_0}
\left(
\frac{x_0^2}{\lambda t}
\right)^{\frac{1}{2}+\frac{\lambda}{2\Delta}}\times
\nonumber
\\
&&\frac
{
(1-e^{-2u})^{\frac{1}{2}+\frac{\lambda}{2\Delta}}
}
{
(\sinh 2u)^{\frac{1}{2}}
}
\exp\left(-\frac{F(x,t)}{2\Delta t}\right)~,
\label{asymp2}
\end{eqnarray}
where $F(x,t)=x^2+x_0^2-\lambda t\coth u$ or by insertion
\begin{eqnarray}
F(x,t)=x^2+x_0^2-[(2xx_0)^2+(\lambda t)^2]^{1/2}~.
\end{eqnarray}

In the short time limit $\lambda t\ll 2xx_0$  $u$ is small and 
$\sinh u\sim u$, i.e., $u\sim \lambda t/2xx_0$ and we
obtain $P(x,t)\sim (1/2\pi\Delta t)^{1/2}\exp(-(x-x_0)^2/2\Delta t)$,
yielding $\delta(x-x_0)$ for $t=0$.
The weak coupling limit $\lambda\rightarrow 0$
for fixed $x$ and $t$ is also consistent; we obtain 
$P(x,t)\sim (1/2\pi\Delta t)^{1/2}\exp(-(x-x_0)^2/2\Delta t)$.
For weak noise the peak of the distribution is determined by
the condition $F=0$, yielding
$x=(x_0^2-\lambda t)^{1/2}$ and the peak thus follows
the noiseless finite-time-singularity solution (\ref{sollog}).
\bibliography{articles,books}

\end{document}